\documentclass[amsmath,amssymb,reprint,groupedaddress,showpacs,aps,prl]{revtex4-1}
\usepackage{graphicx}% Include figure files
\usepackage{color}
\usepackage{caption}
\definecolor{red}{rgb}{1,0.,0}

\newcommand{\ket}[1]{\left| #1 \right>} % for Dirac bras

\def\smath#1{\text{\scalebox{.8}{$#1$}}}
\begin{document}

%%% TITLE et al. %%%%
\title{Superdense coding over optical fiber links with complete Bell-state measurements}

\author{Brian P. Williams}
\affiliation{Quantum Computing Institute, Oak Ridge National Laboratory, Oak Ridge, Tennessee USA 37831}
\email{williamsbp@ornl.gov}
\author{Ronald J. Sadlier}
\affiliation{Quantum Computing Institute, Oak Ridge National Laboratory, Oak Ridge, Tennessee USA 37831}
\email{sadlierrj@ornl.gov}
\author{Travis S. Humble}
\affiliation{Quantum Computing Institute, Oak Ridge National Laboratory, Oak Ridge, Tennessee USA 37831}
\email{humblets@ornl.gov}

\begin{abstract}
Adopting quantum communication to modern networking requires transmitting quantum information through fiber-based infrastructure. We report the first demonstration of superdense coding over optical fiber links, taking advantage of a complete Bell-state measurement enabled by time-polarization hyperentanglement, linear optics, and common single-photon detectors. We demonstrate the highest single-qubit channel capacity to date utilizing linear optics, $1.665 \pm 0.018$, and we provide a full experimental implementation of a hybrid, quantum-classical communication protocol for image transfer. 
\end{abstract}
\maketitle
Superdense coding enables one qubit to carry two bits of information between a sender Bob and receiver Alice when they share an entanglement resource, perhaps distributed at ``off-peak" times \cite{densecoding92}. Bob can choose to use this quantum ability at a time of optimal advantage, after the time of distribution, when he and Alice have access to quantum memory \cite{lvovsky2009optical}. A significant challenge in realizing superdense coding is the need to perform a complete Bell-state measurement (BSM) on the photon pair, which is not possible using only linear optics and a single degree of shared entanglement \cite{lutkenhaus1999bell,Vaidman99}. While nonlinear optics \cite{kim2001quantum} or utilization of ancillary photons and linear optics \cite{warren} enable a complete BSM, these methods are challenged by inefficiency and impracticality. However, a complete BSM with linear optics can be performed by using entanglement in additional degrees of freedom, so-called hyperentanglement \cite{kwiat1997hyper,nick}. Complete BSM implementations have been demonstrated previously using states hyperentangled in orbital angular momentum and polarization degrees of freedom \cite{barreiro2008beating} as well as with states hyperentangled in momentum and polarization \cite{barbieri2007complete}. However, these states are not compatible with transmission through fiber-based networks, which form the backbone of modern telecommunication systems. Previously, Schuck et al.~have shown that time-polarization hyperentanglement can be used for complete and deterministic BSM \cite{Schuck2006complete}, but required number resolving detectors to identify the states completely. Yet this encoding is attractive in that it permits efficient transmission through optical fiber and that some photon pair sources generate time entanglement for ``free". 
\begin{figure}[tbh]
\centering
\captionsetup{justification=RaggedRight}
\includegraphics[width=0.9\linewidth]{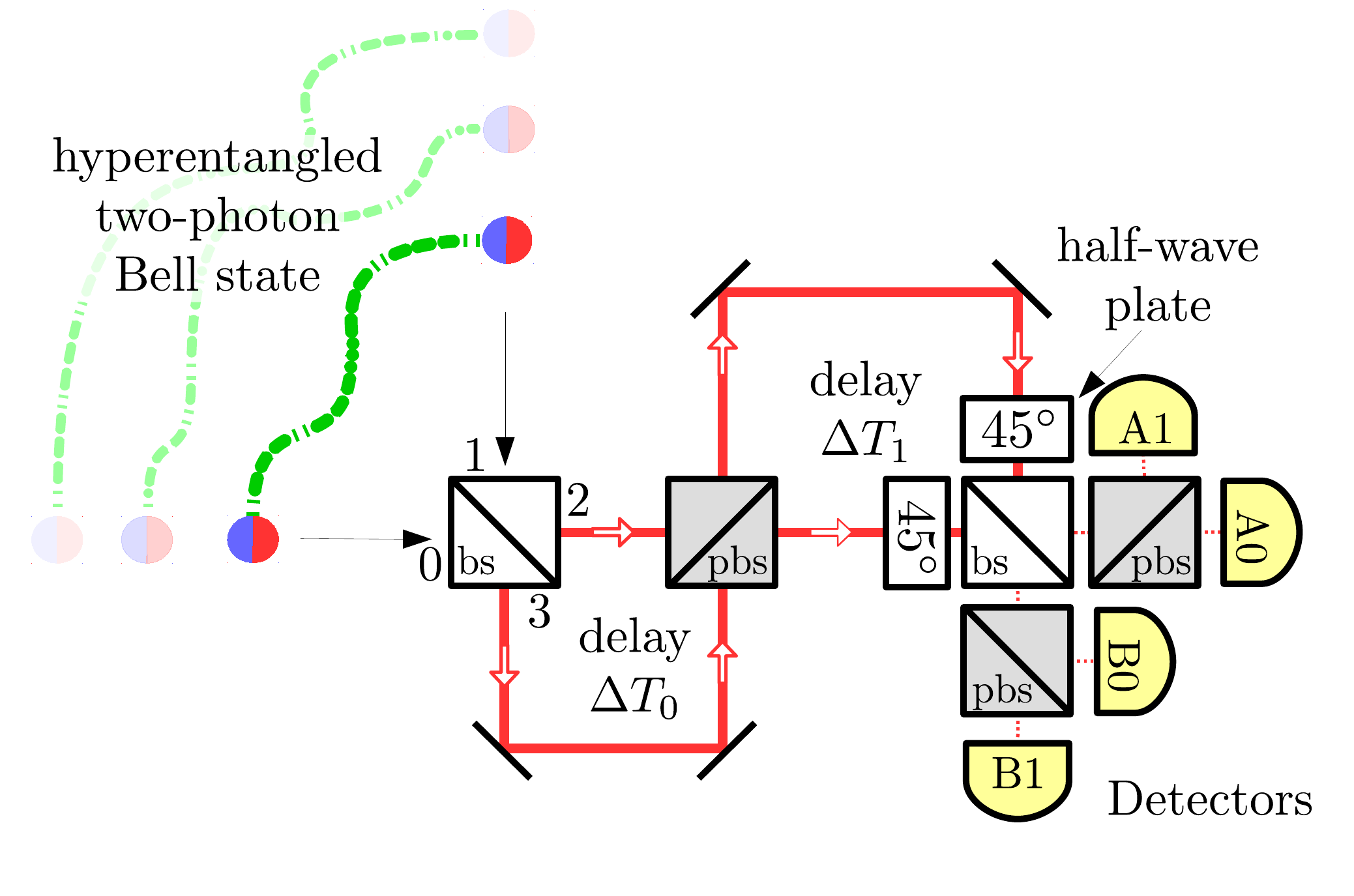}
\caption{Our complete Bell state measurement is achieved using linear optics and time-polarization hyperentanglement. Two-photon  interference is orchestrated to ensure Bell state detection signatures do not involve two-photons coincident on a single detector. bs$\equiv$beamsplitter, pbs$\equiv$polarizing beamsplitter}
\label{fig1}
\end{figure}
\par
In this article, we report results from an experimental demonstration of superdense coding over optical fiber links using a complete BSM based on time-polarization hyperentanglement. Our novel implementation requires only linear optics and common single-photon detectors. The resulting combination of a hyperentangled photon pair source and novel discrimination device yields an observed channel capacity of 1.665$\pm 0.018$, the highest reported to date for a single-qubit and linear optics. We demonstrate the feasibility of this setup by transmitting a 3.4 kB image that is recovered with $87\%$ fidelity. This proof-of-principle milestone opens up the potential integration of practical superdense coding within modern fiber-based telecommunication infrastructure \cite{quantumNetworking}.
\par
\begin{figure}[tbh]
\centering
\captionsetup{justification=RaggedRight}
\includegraphics[width=0.9\linewidth]{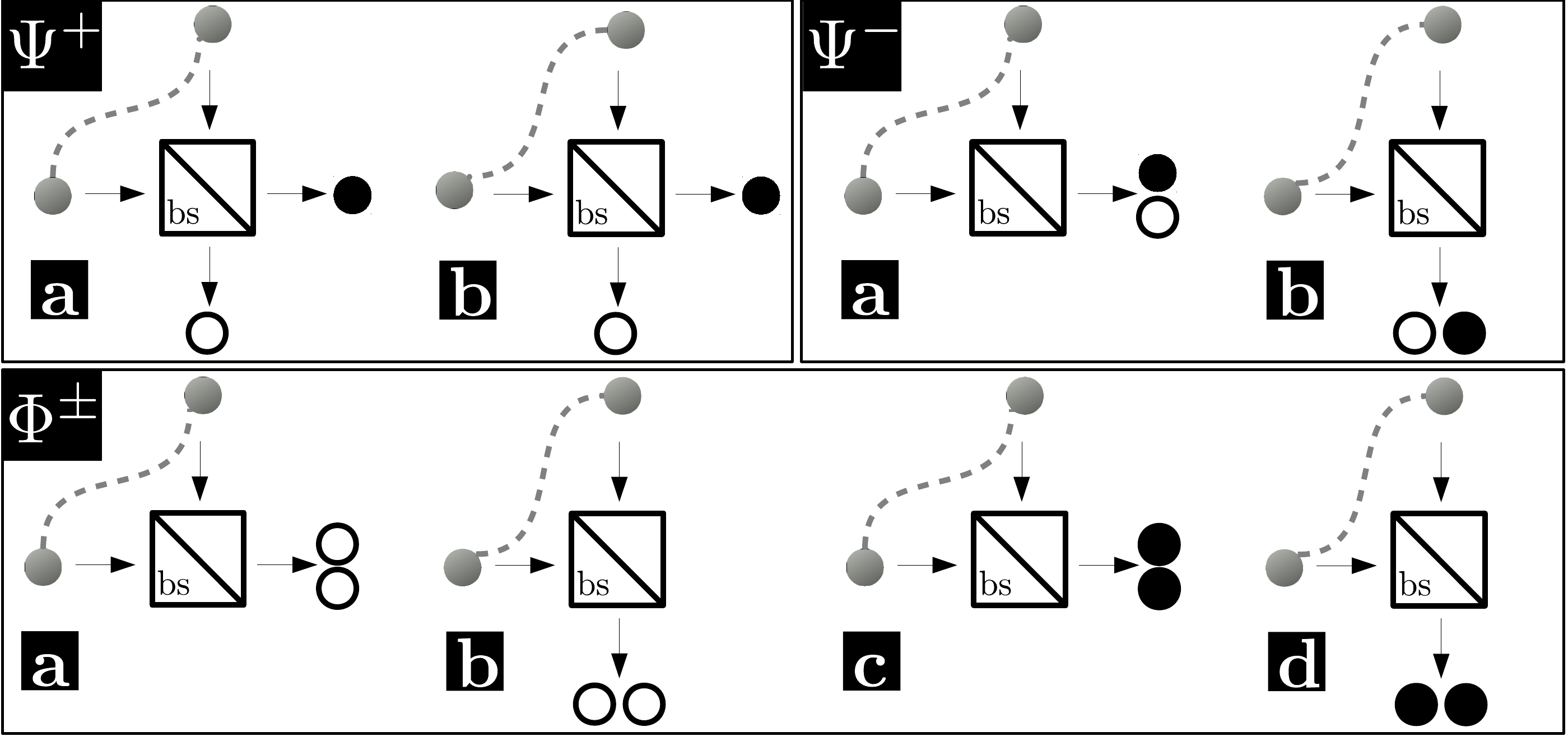}
\caption{Hong-Ou-Mandel interference at a beam splitter leads to unique outputs for the four Bell states. This leads to partial Bell state discrimination, perfectly identifying $\Psi^+$ and  $\Psi^-$ from $\Phi$ states. The $\Phi^+$ and $\Phi^-$ states are left  indistinguishable. }
\label{fig2}
\end{figure}

We use the experimental configuration for the complete BSM shown in Fig.~\ref{fig1}. The principle of operation behind the BSM device is to use the temporal and polarization coherences of the photon pair to produce unique detection outcomes for each of the four polarization-encoded Bell states,
\begin{align}
 \left|\Phi^\pm\right\rangle&=(\left|H_0 H_1\right\rangle\pm\left|V_0 V_1\right\rangle)/\sqrt{2} \label{phi}\\ 
 \left|\Psi^\pm\right\rangle&=(\left|H_0 V_1\right\rangle\pm\left|V_0 H_1\right\rangle)/\sqrt{2}
 \label{psi}
\end{align}
with $H$ and $V$ denoting horizontal and vertical polarization, respectively, and the subscripts $0$ and $1$ denote the orthogonal spatial modes. The entangled photons enter the interferometer in Fig.~\ref{fig1} and output to modes $2$ and $3$. As shown in Fig.~\ref{fig2}, Hong-Ou-Mandel interference at this first beam splitter leads to a unique output for each of the four Bell state inputs. Schuck et al.~used a similar arrangement followed by photon number resolving detection to identify the output state. 
\par
We avoid the need for photon number resolving detectors by orchestrating the two-photon interference such our unique BSM signatures do not involve the photon pair coincident at a detector--that would require a number resolving detector to observe deterministically. This orchestration results in the four Bell states having three distinct photon pair detection times.  As shown by Figs.~\ref{fig1} and \ref{fig2}, $\Phi^\pm$ states result in photon pair detections with no delay between photon time of arrival. This is necessary since the photons in that state copropagate through the interferometer. By comparison, the $\Psi^+$ state results in a relative time delay of $\Delta T_0$ between members of the photon pair. This is due to the members taking separate paths in the first interferometer segment and their copropagation in the second segment.   A similarly unique signature for the $\Psi^-$ state arises when those photons copropagate in the first segment and take separate paths in the second segment of the BSM. This requires the $\Psi^-$ photon pair to have a relative time delay of $\Delta T_1$.
\par
The distinct propagations for each Bell state illustrate that our apparatus is a ``4-in-1" two-photon interferometer that utilizes elements from several fundamental quantum information ideas including Hong-Ou-Mandel interference \cite{HOM}, partial Bell state analysis \cite{weinfurter1994}, and time-entanglement \cite{franson1989bell}. Unlike previous experiments that discard a portion of the quantum coherence by segregating outputs after the first beamsplitter \cite{Schuck2006complete}, we retain all coherence present to assist in the complete Bell state discrimination. In conjunction with the aforementioned time signatures, the path delayed two-photon interferometers ensure unique fully determinable detection outcomes that do not include two photons with the same polarization in the same output port, a limitation of a previous experiment \cite{Schuck2006complete}.
\par
We verify the design of the unique detection capabilities for this device by considering the symmetric beamsplitter relations \cite{Schuck2006complete,ou2007multi},
\begin{align}
&\ket{H_0} \!\overset{BS}{\rightarrow}\! \left(\ket{H_2}+i \ket{H_3}\right)/{\smath{ \sqrt{2}}}\quad \ket{H_1} \!\overset{BS}{\rightarrow}\!\left(i \ket{H_2}+\ket{H_3} \right)/{\smath{ \sqrt{2}}}\nonumber\\
&\ket{V_0}\! \overset{BS}{\rightarrow}\! \left(\ket{V_2 \;}-i \ket{V_3 \;}\right)/{\smath{ \sqrt{2}}}\quad \ket{V_1 \;} \!\overset{BS}{\rightarrow}\!\left(-i \ket{V_2}+ \ket{V_3}\right)/{\smath{ \sqrt{2}}}\nonumber
\end{align}
where the ports $0, 1, 2,$ and $3$ are identified at the first beamsplitter in Fig. \ref{fig1}. Additionally, the effect of a  $45^\circ$ polarization rotation on a single-photon using a half-wave plate, a Hadamard gate, is modeled as
\begin{align}
\ket{H} \overset{45^\circ}{\rightarrow}\left( \ket{H}+\ket{V} \right)/{\smath{ \sqrt{2}}} \quad\quad \ket{V} \overset{45^\circ}{\rightarrow}\left( \ket{H}-\ket{V} \right)/{\smath{ \sqrt{2}}}
\textrm{.}\nonumber
\end{align}
Given these relations, one may verify that with phases $\omega \Delta T_0=2n\pi$ and $\omega \Delta T_1=2m\pi$, $n$ and $m$ are integers, the output states for the four Bell states given by Eqs.~(\ref{phi}) and (\ref{psi}) yield the interferometer output states
\begin{align}\left|\Phi^+\right\rangle& \!\rightarrow\! (\ket{H_A V_A}+\ket{H_B V_B})/{\smath{ \sqrt{2}}}\nonumber \\ 
\left|\Phi^-\right\rangle& \!\rightarrow\! (\ket{H_A H_B}-\ket{V_A V_B})/{\smath{ \sqrt{2}}}\nonumber \\
\left|\Psi^+\right\rangle& \!\rightarrow\! \left(\ket{H_A H_B'}+\ket{H_A' H_B}+\ket{V_A V_B'}+\ket{V_A' V_B}\right.\nonumber\\
&\quad+\left.\ket{H_A V_B'}-\ket{H_A' V_B}+\ket{V_A H_B'}-\ket{V_A' H_B}\right)/{\smath{ \sqrt{8}}} \nonumber \\
\left|\Psi^-\right\rangle& \!\rightarrow\! \left(\ket{H_A V_A''}+\ket{H_A'' V_A}-\ket{H_B V_B''}-\ket{H_B'' V_B}\right.\nonumber\\
&\quad\!+\!\left.i\left[\;\ket{H_A V_B''}\!-\!\ket{H_A'' V_B}\!+\!\ket{V_A H_B''}\!-\!\ket{V_A'' H_B}\right] \right)/{\smath{ \sqrt{8}}}
\nonumber\end{align}
where $A$ and $B$ label the output ports and prime and double prime indicates a $\Delta T_0$ and $\Delta T_1$ delayed photon, respectively. This set of outputs is smaller than what would be observed for photons separable in the temporal degree of freedom, i.e., photons that are not hyperentangled. A time-entangled photon pair can take one of multiple paths, 2 to 4 depending on the state, but these paths are indeterminable--they interfere upon leaving the interferometer. This interference and subsequent interferometer phase sensitivity allow us to choose the deterministic and measurable Bell state detection signatures given in Table I. In addition to superdense coding, this novel BSM may serve as an entanglement witness for fiber-based quantum seal applications \cite{tamperSeal}. 
\begin{figure}[tbh]
\centering
\captionsetup{justification=RaggedRight}
\begin{minipage}[c]{\linewidth}
\captionof{table}{Bell state detection signatures include a time $\Delta t$ between photon detections, same or different port detection, and same ($\parallel$) or different ($\perp$) polarizations.}
\centering
\vspace{-5pt}
\begin{tabular}{|c|c|}
\hline
\phantom{\huge{O}}\!\!\!\!\!\!\!\!$\Phi^+$ & $\Delta t = 0$, $\parallel$ ports, $\perp$ polarization \\ \hline
\phantom{\huge{O}}\!\!\!\!\!\!\!\!$\Phi^-$ & $\Delta t = 0$, $\perp$ ports, $\parallel$ polarization \\ \hline
\phantom{\huge{O}}\!\!\!\!\!\!\!\!$\Psi^+$ & $\Delta t = \Delta T_0$, $\perp$ ports \\ \hline
\phantom{\huge{O}}\!\!\!\!\!\!\!\!$\Psi^-$ & $\Delta t = \Delta T_1$, $\perp$ polarization \\ \hline
\end{tabular}
\end{minipage}
\end{figure}
\begin{figure*}[tbh]
\centering
\captionsetup{justification=RaggedRight}
\includegraphics[width=0.6\linewidth]{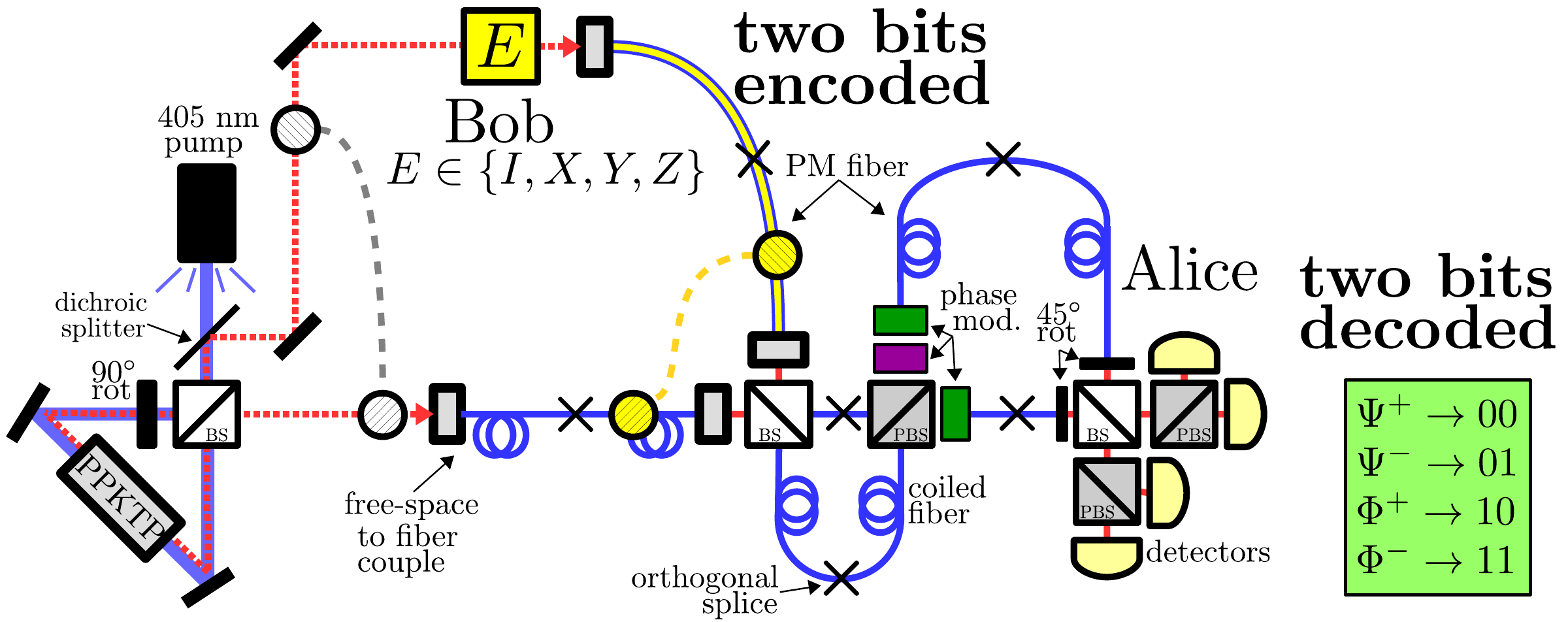}
\caption{To implement superdense coding, Alice and Bob initially each receive one photon from a a time-polarization hyperentangled photon pair. Bob performs one of four operations on his photon which encodes two bits on the nonlocal two-photon Bell state. Bob transmits his photon to Alice who performs a Bell state measurement, i.e. decodes two bits. bs$\equiv$beamsplitter, pbs$\equiv$polarizing beamsplitter, PPKTP$\equiv$potassium titanyl phosphate\label{experiment}}
\end{figure*}
\par
Our experimental implementation utilized an $810$ nm entangled photon pair source characterized to produce each Bell state with 0.97 fidelity. The pair source is based on SPDC using a potassium titanyl phosphate (PPKTP) nonlinear crystal within a Sagnac loop to produce a polarization-entangled biphoton state \cite{WongSagnac}. The temporal entanglement results from the long coherence time of the continuous-wave 405 nm diode laser pump. Pumping with $1.85$ mW produced $2.27 \times 10^5$ pairs/second. Bob encodes each of the four Bell states utilizing two liquid crystal waveplates. The first is oriented to apply an $I$=$\left(\begin{smallmatrix}1&0\\0&1\end{smallmatrix}\right)$ or $X$=$\left(\begin{smallmatrix}0&1\\1&0\end{smallmatrix}\right)$ gate and the second applies an $I$ or $Z$=$\left(\begin{smallmatrix}1&0\\0&\textrm{-}1\end{smallmatrix}\right)$ gate. Each combination $II$, $IZ$, $XI$, $XZ$ encodes one of the four Bell states. Due to losses, the coincident count rate observed by Alice was approximately 200 cps.
\par
Given our timing resolution of 4 ns we required the interferometer to have temporal delays $\Delta T_0\approx 5$ ns and $\Delta T_1\approx 10$ ns. Due to these large time delays, we chose to implement the experimental apparatus in polarization maintaining (PM) optical fiber as seen in Fig. \ref{experiment}. The fiber lengths needed for the first loop were 1 and 2 meters, for the short and long paths, respectively, and, similarly, 2 and 4 meters in the second loop. Each fiber path consists of 2 fibers of equal length orthogonally connected such that the macroscopic temporal effects of birefringence were compensated. Experimentally, we reconfigured a simple fiber connector such that the axes were aligned orthogonally. These connections are indicated in Fig. \ref{experiment} as a $\times$ symbol between two optical fibers. The use of PM optical fiber requires that we take into account the microscopic difference in phase due to birefringence in different paths. This phase compensation and the calibration of the phases needed to ensure unique BSM signatures is accomplished using liquid crystal waveplates. The phase of each photon polarization is individually modulated in the long path of the second loop and the phase of one photon polarization is modulated in the short path of the second loop. Due to the long path lengths, thermal drift of the fibers and frequency drift of the pump laser limits stable operation to approximately 100 seconds before recalibration is necessary. 
\begin{figure}[tbh]
\begin{minipage}[c]{\linewidth}
\centering
\captionof{table}{Raw 5 second counts of Bell state $y$ \\ received given an sent state $x$.}\vspace{-10pt}
\begin{tabular}{|c|c|c|c|c|}\hline
$_x \! \diagdown \! ^y$ &$\Phi^-$& $\Phi^+$& $\Psi^-$ & $\Phi^+$\\\hline
$\Phi^-$ &710 & 8 & 8 & 4\\\hline
$\Phi^+$ &7 & 715 & 9 & 13\\\hline
\end{tabular}
\quad
\begin{tabular}{|c|c|c|c|c|}\hline
$_x \! \diagdown \! ^y$ &$\Phi^-$& $\Phi^+$& $\Psi^-$ & $\Phi^+$\\\hline
$\Psi^-$ &15 & 8 & 748 & 9\\\hline
$\Psi^+$ &34 & 23 & 15 & 840\\\hline
\end{tabular}
\end{minipage}
\captionsetup{justification=RaggedRight}
\includegraphics[width=0.8\linewidth]{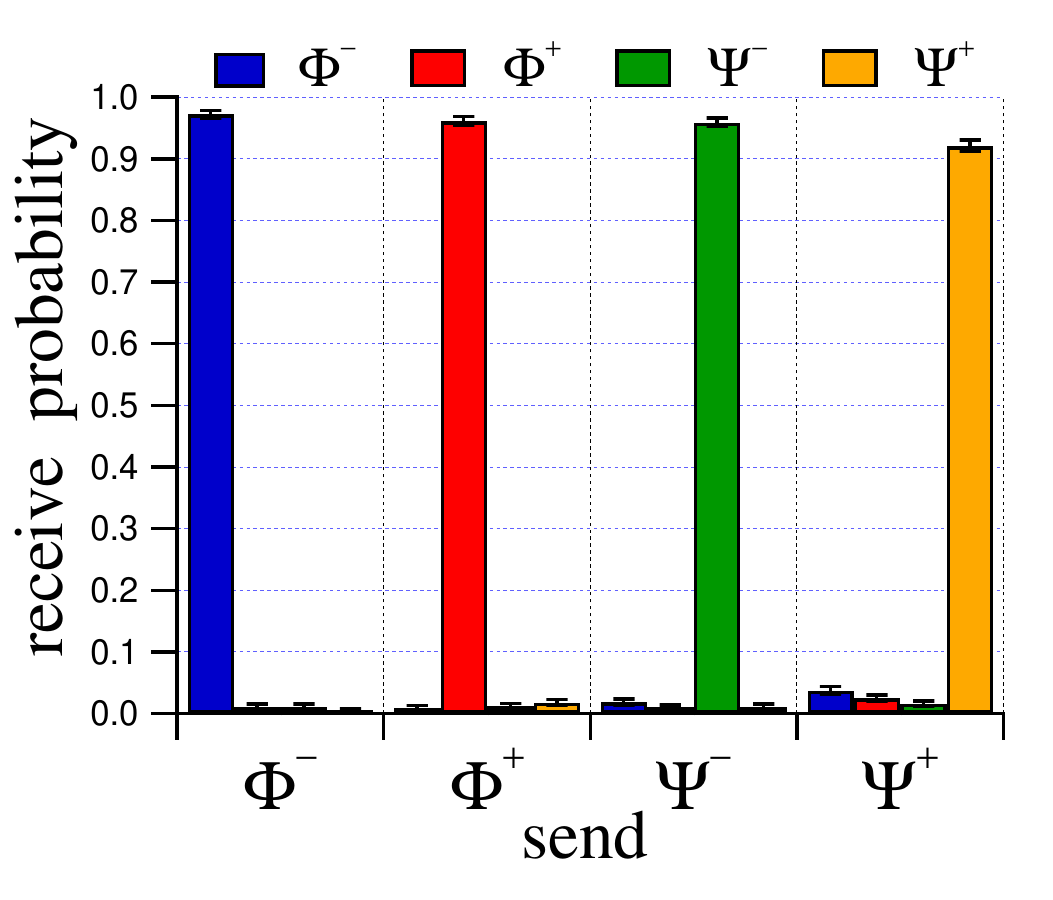}
\caption{The conditional probabilities $p(y|x)$ of receiving state $y$ given the sent state $x$ using our apparatus gives the highest channel capacity to date, 1.665$\pm$0.018, utilizing a single-qubit and linear optics.\label{barplot}}
\end{figure}

We characterized the BSM performance by calibrating the interferometer to optimal discrimination settings, preparing each of the four Bell states, and recording the state detected by the apparatus. In estimating the success probabilities, we recorded approximately 800 detection signatures for each state, which occurred over a 5 second interval. The raw data collected is given in Table II.  The maximum-likelihood estimate for the conditional probabilities with accidental coincidences included are given in Fig. \ref{barplot}.

The channel capacity $C$ \cite{coverThomas} is the maximum mutual information $I(x;y)$ possible for a given set of conditional probabilities $p(y|x)$ where $x$ is the state prepared by Bob and $y$ is the state decoded by Bob. This capacity is 
\begin{equation}C=\overset{\textrm{{\large max}}}{p(\mathbf{x})} \sum_{y=0}^{3}\sum_{x=0}^{3} p(y|x)p(x) \log\left(\frac{p(y|x)}{\sum_{x'=0}^{3}p(y|x')p(x')}\right)\nonumber\end{equation}
where  $p(\mathbf{x})=\{p(\Phi^-),p(\Phi^+),p(\Psi^-),p(\Psi^+)\}$ maximizes the capacity. For the conditional probabilities given in Fig. \ref{barplot} we find the channel capacity to be 1.665$\pm$0.018 for arguments $p(\Phi^-)$=$0.262$, $p(\Phi^+)$=$0.256$, $p(\Psi^-)$=$0.256$, and $p(\Phi^+)$=$0.226$. The maximum channel capacity possible using linear optics, a single-qubit, and a single degree of entanglement is 1.585 \footnote{We note that methods have been proposed to achieve a channel capacity of 2 utilizing linear optics and one photon on average \cite{pavel}.}. Our experiment has exceeded this bound as well as the previous highest channel capacity  reported for a single-qubit using linear optics \cite{barreiro2008beating}.
\par
We next demonstrate the use of superdense coding as part of a hybrid, quantum-classical protocol for communicating between users Alice and Bob. We make use of conventional optical network communication between two users to coordinate Alice's quantum transmitter and Bob's quantum receiver \cite{Humble2014,Sadlier2016}. The basic steps in the protocol for this demonstration are:
\begin{itemize}\vspace{-5pt}
\item[1.] Bob transmits a classical send request to Alice.\vspace{-5pt}
\item[2.] Alice returns a classical acknowledgement to Bob.\vspace{-5pt}
\item[3.] Bob transmits two bits using superdense coding.\vspace{-5pt}
\item[4.] Alice receives and decodes the two-bit message.\vspace{-5pt}
\item[5.] Alice transmits a classical receipt to Bob.\vspace{-5pt}
\end{itemize}
This protocol is repeated $n$ times until Bob has completed transmission of a $2n$-bit message to Alice. Due to the spontaneous nature of the photon pair source, only the first photon-pair detected during step 3 above is used even though other photon pairs may be detected. In this regard, the classical messages before and after a burst of superdense coding act to frame each two-bit message. While this incurs a large  overhead, it ensures the detected bits are properly framed so that the transmitted message can be reliably constructed. A less conservative protocol would permit Bob to transmit multiple two-bit messages uninterrupted and require Alice to properly partition this sequence into the corresponding frames.

\begin{figure}[t]
\captionsetup{justification=RaggedRight,width=\linewidth}
\begin{minipage}[t]{0.4\linewidth}
\includegraphics[width=\linewidth]{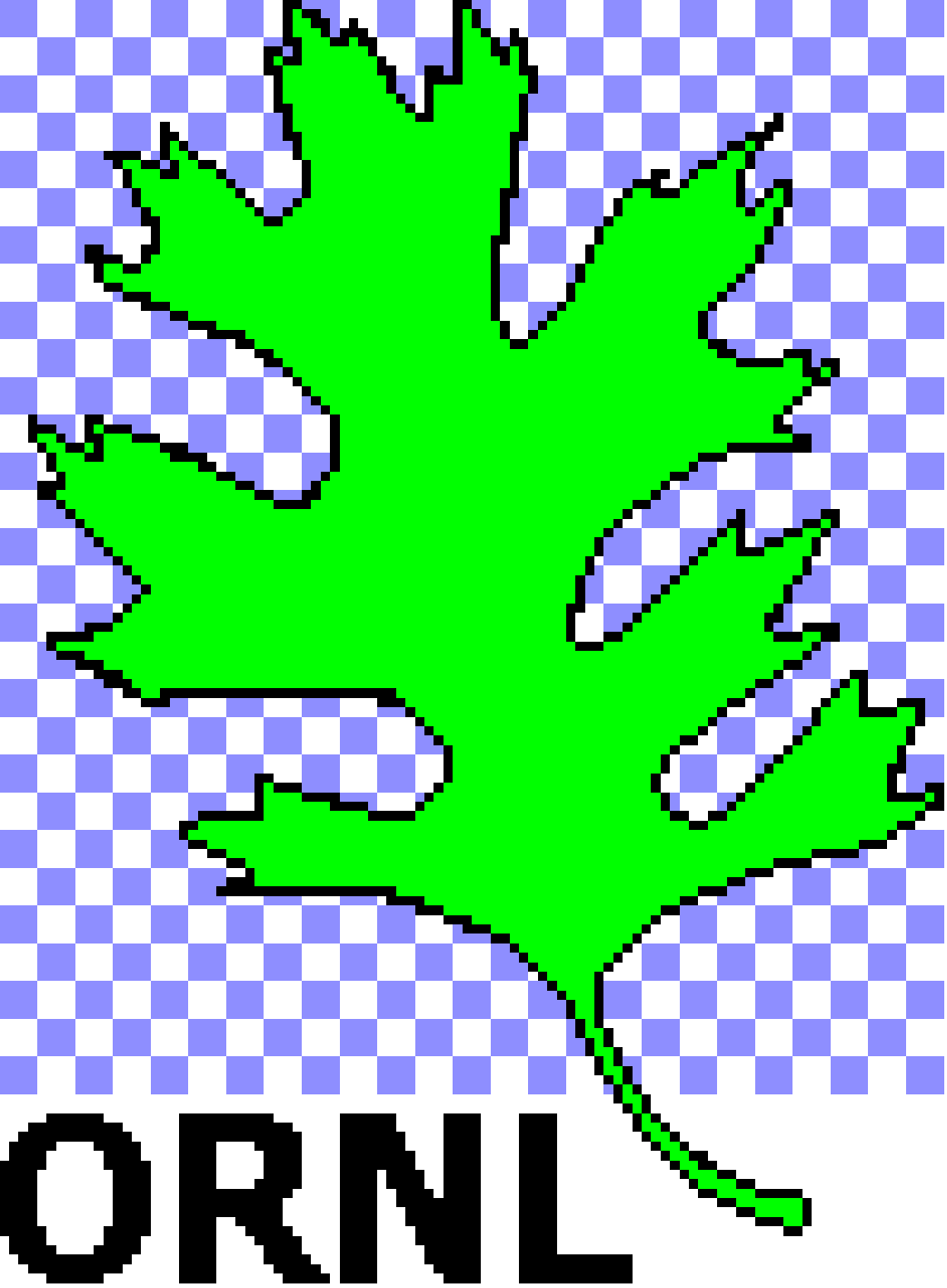}
\end{minipage}
\quad
\begin{minipage}[t]{0.4\linewidth}
\includegraphics[width=\linewidth]{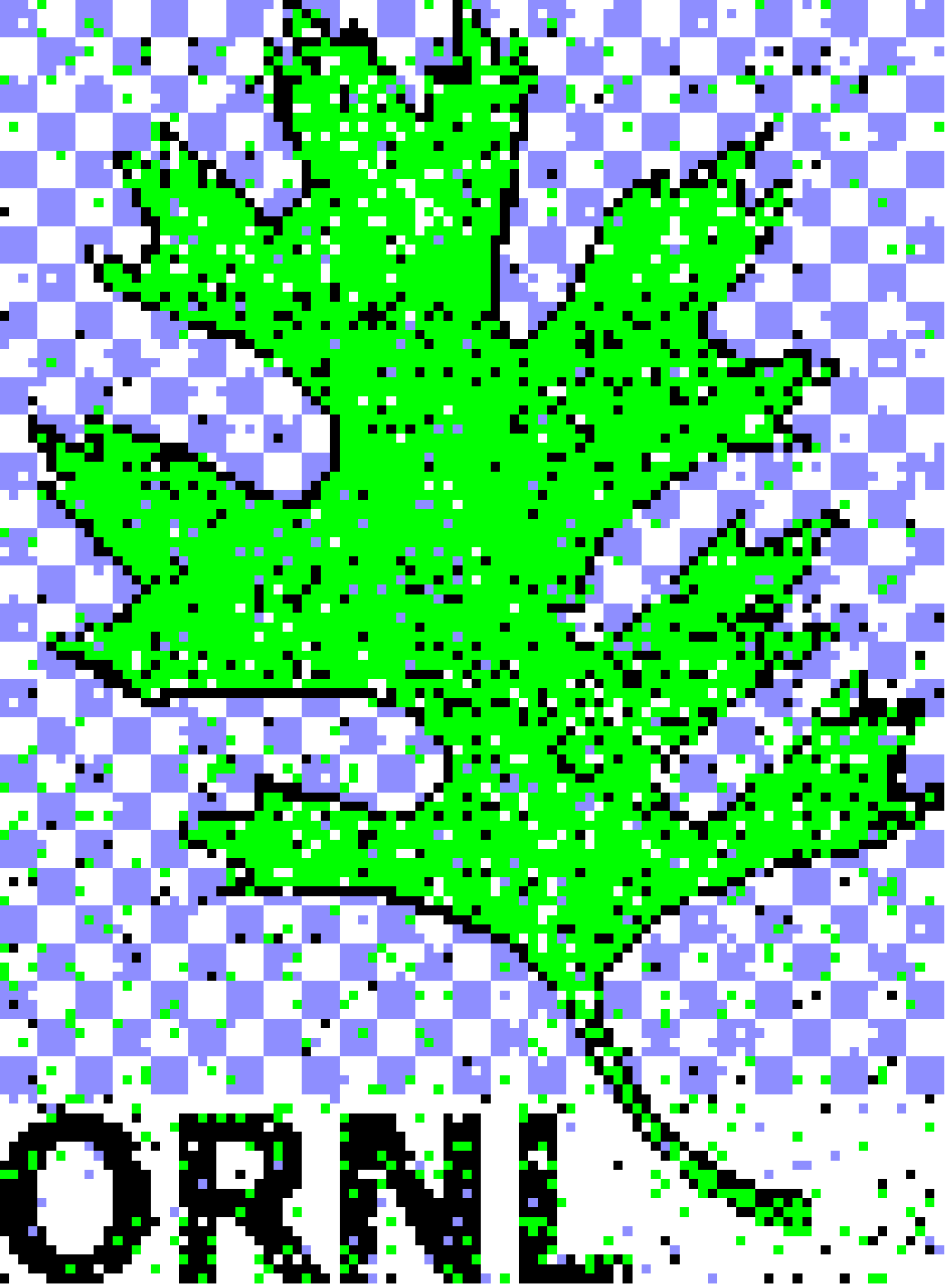}
\end{minipage}
\caption{ (Left): The original 4-color 100x136 pixel 3.4kB image. (Right): The image received using superdense coding. The calculated fidelity was 87$\%$.\label{leaf}}
\end{figure}

As a demonstration of the system capabilities, we have transmitted the 3.4 kB image shown in Fig. \ref{leaf}(left).  The corresponding received image  is shown in Fig. \ref{leaf}(right). The received image is calculated to have a fidelity of $87\%$. The errors in the received image result from drift in the interferometer during transmission, phase miscalibration, and imperfect state generation. It should be apparent that the fidelity of the received image is less than that of the optimal characterization presented in Fig.~\ref{barplot}. This is due to performance trade-offs in system operation that reduce the collection time at the expense of reduced detection fidelity.  For the current demonstration the effective bit rate during operation that included a classical send message, phase encoding using liquid crystal waveplates, a classical receive message, and periodic calibration was 1 bit/sec.  The fiber link for this system was 2 meters of PM optical fiber. We do not foresee any fundamental impediments to using longer optical fibers beyond increased losses and cost. In particular, an implementation using standard single-mode fiber could operate by using a polarization correction or stabilization system. 
\par
Our experiment could be improved by reducing the spontaneity of the source by utilizing a pulsed pump laser, in which case the temporal delays in the interferometer would need to be multiples of the pulse period $t_p$, for instance $\Delta T_0=t_p$ and $\Delta T_1=2 t_p$. Achieving a finer time resolution would allow for smaller delays $\Delta T_0$ and $\Delta T_1$ which would improve the stability, i.e. the drift would be reduced. Packaging the device or using integrated optics would substantially increase the bit-rate by reducing the photon pair loss due to free-space to fiber coupling inefficiencies. The speed of Bob's encoding is limited by the response time of the liquid crystal waveplates which is on the order of milliseconds. Faster options include lithium-niobate phase modulators \cite{LiNbO3} which can achieve modulation speeds on the order of nanoseconds.

In conclusion, we have achieved a single-qubit channel capacity of $1.665 \pm 0.018$ over optical fiber links enabled by a complete Bell state measurement using linear optics, hyperentanglement, and common single-photon detectors. This channel capacity is the highest to date for a single-qubit and linear optics. Our novel interferometric design allows ``off-the-shelf" single-photon detectors to enable the complete Bell state discrimination instead of the number-resolving detectors required by previous experiments \cite{Schuck2006complete}. To our knowledge, this is the first demonstration of superdense coding over optical fiber and a step towards practical realization of superdense coding. Alongside our demonstration of a hybrid quantum-classical transfer protocol, these results represents a step toward the future integration of quantum communication with fiber-based networks \cite{Humble2013,Dasari2015}.

This work was supported by the United States Army Research Laboratory. This manuscript has been authored by UT-Battelle, LLC under Contract No. DE-AC05-00OR22725 with the U.S. Department of Energy. The United States Government retains and the publisher, by accepting the article for publication, acknowledges that the United States Government retains a non-exclusive, paid-up, irrevocable, worldwide license to publish or reproduce the published form of this manuscript, or allow others to do so, for United States Government purposes. The Department of Energy will provide public access to these results of federally sponsored research in accordance with the DOE Public Access Plan (http://energy.gov/downloads/doe-public-access-plan).

\bibliography{CDBSM.bib}
\end{document}